\documentstyle[sprocl,epsfig]{article}

\def\Journal#1#2#3#4{{#1} {\bf #2}, #3 (#4)}

\def\NPB{{\em Nucl. Phys.} B}
\def\PLB{{\em Phys. Lett.}  B}
\def\PRL{\em Phys. Rev. Lett.}
\def\PRD{{\em Phys. Rev.} D}

\newcommand{\req}[1]{(\ref{#1})}
\newcommand{\Beq}{\begin{equation}}
\newcommand{\Eeq}{\end{equation}}
\newcommand{\beq}{\begin{displaymath}}
\newcommand{\eeq}{\end{displaymath}}
\newcommand{\Beqa}{\begin{eqnarray}}
\newcommand{\Eeqa}{\end{eqnarray}}
\newcommand{\beqa}{\begin{eqnarray*}}
\newcommand{\eeqa}{\end{eqnarray*}}
\newcommand{\nUV}{\mu_{R}^{2}}
\newcommand{\nIR}{\mu_{F}^{2}}

\newcommand{\x}{\overline{x} \,}
\newcommand{\y}{\overline{y} \,}

\begin{document}

%%% start MY title page %%%%%%%%%%%%%

\begin{titlepage}

\begin{flushright}
IRB-TH-7/99\\
August, 1999
\end{flushright}

\vspace*{2cm}

\begin{center}
\Large\bf On the PQCD prediction for the pion form factor  
\vspace*{0.3truecm}
\end{center}

\vspace{1.2cm}

\begin{center}
\large B. Meli\'{c}, B. Ni\v{z}i\'{c}, K. Passek\\
\vspace*{0.3cm}
\normalsize
{\sl Theoretical Physics Division, Rudjer Bo\v skovi\'c Institute,
P.O.Box 1016, HR-10001 Zagreb, Croatia\\[3pt]
E-mail: {\it melic@thphys.irb.hr, nizic@thphys.irb.hr,
passek@thphys.irb.hr}}
\end{center}

\vspace{1.1cm}

\begin{center}
{\bf Abstract}\\[0.3cm]
\parbox{12cm}
{We comment on the results of a complete leading-twist
next-to-leading order QCD analysis of the 
spacelike pion electromagnetic form factor at large-momentum
transfer $Q$.
For the asymptotic distribution amplitude, 
we have examined the sensitivity of the predictions 
to the choice of the renormalization scale.
The results show that, regarding the size of the radiative
corrections, reliable 
perturbative predictions for the pion electromagnetic form factor 
can already be made at a momentum transfer $Q$ of the order of
$5$ to $10$ GeV. 
}
\end{center}

\vspace{1.5cm}

\begin{center}
{\sl Talk given by K. Passek at the \\
Joint INT/ Jefferson Lab Workshop on\\
Exclusive \& Semiexclusive
Processes at High Momentum Transfer \\
%Newport News, Virginia\\
20-22 May 1999 \\
To appear in the Proceedings}
\end{center}

\end{titlepage}
\thispagestyle{empty}
\vbox{}
\newpage

\setcounter{page}{1}

%%% end MY title page %%%%%%%%%%%%%
\title{ON THE PQCD PREDICTION FOR THE PION FORM FACTOR}

\author{B. MELI\'{C}, B. NI\v{Z}I\'{C}, K. PASSEK}

\address{Theoretical Physics Division, 
        Rudjer Bo\v{s}kovi\'{c} Institute, 
        P.O. Box 1016, HR-10001 Zagreb, Croatia\\
        E-mail: melic@thphys.irb.hr, nizic@thphys.irb.hr, passek@thphys.irb.hr}

\maketitle\abstracts{ 
We comment on the results of a complete leading-twist
next-to-leading order QCD analysis of the 
spacelike pion electromagnetic form factor at large-momentum
transfer $Q$.
For the asymptotic distribution amplitude, 
we have examined the sensitivity of the predictions 
to the choice of the renormalization scale.
The results show that, regarding the size of the radiative
corrections, reliable 
perturbative predictions for the pion electromagnetic form factor 
can already be made at a momentum transfer $Q$ of the order of
$5$ to $10$ GeV. 
}

\section{Introduction}

The study of exclusive processes at large-momentum transfer 
represents
a challenging area for application of perturbative QCD (PQCD).
Although the PQCD approach \cite{sHSA}
undoubtedly represents an adequate and efficient tool for analyzing
exclusive processes at very large momentum transfer,
its applicability to these processes at experimentally accessible 
momentum transfer has been long debated and attracted much attention.
In a moderate energy region (a few GeV) 
soft contributions 
could still be substantial.\cite{soft} 
Further on, the self-consistency of the PQCD  approach was questioned
regarding the nonfactorizing end-point contributions.\cite{IsLS84etc} 
It has been shown, that the incorporation of 
the Sudakov suppression
in the so-called modified hard-scattering approach (mHSA) \cite{LiS92etc,JaK93} 
effectively eliminates soft contributions from the end-point regions and 
that the PQCD approach 
to the pion form factor begins to be self-consistent for a momentum transfer
of about  $Q^2 > 4$ GeV$^2$. 
However, in the PQCD approach to exclusive processes one still has to check
its self-consistency by studying radiative corrections.
It is well known that, unlike in QED, the 
leading-order (LO) predictions in PQCD 
do not have much predictive power, and that higher-order corrections
are important. 
They have a stabilizing effect reducing
the dependence of the predictions on the schemes 
and scales. 

In our recent papers \cite{ourps} we have clarified 
some discrepancies between
previous calculations,\cite{pioneff} 
and by including the complete closed form 
for the NLO evolution of the pion distribution amplitude (DA) 
derived recently,\cite{Mu94etc} 
we have obtained the complete NLO PQCD prediction
for the pion electromagnetic form factor
(within the so-called standard hard-scattering approach (sHSA) 
and using different candidate DAs). 
The size of the NLO correction as well as the size of
the expansion parameter, i.e. QCD running coupling
constant, serve as sensible measures of the self-consistency
of PQCD prediction. But, as the truncation of the
perturbative expansion for the pion form factor 
at finite order causes the residual dependence of the prediction
on the choice of the renormalization and factorization scales
(as well as on the renormalization scheme),
the choices for these scales represent the major
ambiguity in the interpretation of the results.

In this paper we would like to outline 
our calculation \cite{ourps} 
and to address the scale ambiguity problem.

\section{Pion electromagnetic form factor in the sHSA}

In leading twist, the pion electromagnetic form factor
can be expressed by a convolution formula 
\Beq
    F_{\pi}(Q^{2})=\int_{0}^{1} dx \int_{0}^{1} dy \;
                    {\mathit \Phi}^{*}(y,\nIR) \;
                          T_{H}(x,y,Q^{2},\nUV,\nIR) \;
                    {\mathit \Phi}(x,\nIR) \, .
\label{eq:piffcf}
\Eeq

Here $Q^2=-q^2$ is the momentum transfer in the process and
is supposed to be large, 
$\mu_R$ is the renormalization scale,
and $\mu_F$ is the factorization scale at which
soft and hard physics factorize;
$x$ and $y$ ($\x=1-x$ and $\y=1-y$) denote incoming
and outgoing quark (antiquark) momentum fractions. 

The (process dependent)
hard-scattering amplitude $T_{H}(x,y,Q^{2},\nUV,\nIR)$ 
is calculated in 
perturbation theory and represented as a
series in the QCD running coupling constant $\alpha_S(\nUV)$.
We have used the dimensional regularization method and the
$\overline{MS}$ renormalization scheme in our
calculation.\cite{ourps} 

The (process independent)
pion DA ${\mathit\Phi}$ is intrinsically nonperturbative
quantity, 
whose evolution can be calculated perturbatively
and represented as a series in $\alpha_S(\nIR)$.
There are compelling theoretical results \cite{soft,Rad95etc} which disfavor
the end-point concentrated distributions,
and one expects that the pion DA does not differ much
from the asymptotic form.
In this work we comment only on
the results obtained with the asymptotic distribution 
$\phi_{as}(x,\nIR)$. 
%As expected, the NLO evolution of $\phi_{as}(x,\nIR)$ is tiny.

Generally, one can express the NLO form factor as
\Beq
  F_{\pi}(Q^2,\nUV,\nIR) = F_{\pi}^{(0)}(Q^2,\nUV,\nIR)
          + F_{\pi}^{(1)}(Q^2,\nUV,\nIR) \, .
\label{eq:Fpi}
\Eeq
The first term in \req{eq:Fpi} is the LO contribution, while
the second term is the NLO contribution coming
from the NLO correction to the hard-scattering amplitude
as well as arising from the inclusion of the NLO
evolution of the DA. 
For the results obtained using $\phi_{as}(x,\nIR)$ distribution,
the effect of the NLO evolution of the DA is negligible ($\approx 1\%$).

Truncation of the perturbative series of $F_{\pi}(Q^2)$ 
at any finite order causes a residual dependence on the scheme as well as
on the scales (which is explicitly denoted in Eq. \req{eq:Fpi}).
As we approximate $F_{\pi}(Q^2)$ 
only by  two terms of the perturbative series, 
we hope that we can minimize higher-order corrections
by  a suitable choice of $\mu_{R}$ and $\mu_{F}$, 
so that the LO term $F_{\pi}^{(0)}(Q^2,\nUV,\nIR)$
gives a good approximation 
to the complete sum $F_{\pi}(Q^2)$.

We take that a PQCD prediction for pion form factor
can be considered reliable provided 
the corrections to the LO prediction are reasonably small
($< 30\%$) and
the expansion parameter (effective QCD coupling constant) is
acceptably small ($\alpha_S(\mu_R^2)< 0.3$ or $0.5$).
The consistency with the experimental data
is not of much use here since 
reliable experimental data \cite{Be78}
for the pion form factor exist for $Q^2 \leq 4$ GeV$^2$ 
 i.e., 
outside the region in which the perturbative treatment based on
Eq. \req{eq:piffcf} is justified.
It should also be mentioned that 
there are controversial arguments \cite{exp?} 
regarding the reliability of existing experimental data.\cite{Be78}
The new data in this energy region are expected from the CEBAF experiment
E-93-021.

\section{Examining the scale dependence of the NLO prediction}

The simplest and widely used choice for the  $\mu_R$ and $\mu_F$ scales
is
\Beq
      \nUV=\nIR=Q^2 \, ,
\label{eq:nRQ}
\Eeq
the justification for the use of which is mainly pragmatic.
The prediction for the pion form factor
depends very weakly on the choice of the factorization scale $\mu_F$.
Actually, 
taking $\nIR$ to be an effective constant, i.e.,
$\nIR=\left< \nIR \right>$, 
the only $\nIR$ dependence of
the results obtained using $\phi_{as}(x,\nIR)$ distribution
comes from the NLO evolution of the DA 
which is negligible. 

Physically, a more appropriate choice for $\nUV$ 
would be that corresponding to the characteristic
virtualities of the particles in the parton subprocess
(which are considerably lower  than the overall momentum transfer
$Q^2$ i.e., virtuality of the probing photon)
\Beq
     \nUV = a(x,y) \; Q^2 \, . 
\label{eq:axyQ2}
\Eeq
For example, some of 
the physically motivated choices \cite{ourps}  are 
\Beq
    a(x,y) \in
         \{
{\x \y,  \sqrt{\x \y \y} ,
e^{-5/3} \x \y}
         \} \, .
\label{eq:axy}
\Eeq
These correspond, respectively, to the (LO) gluon virtuality,
geometrical mean of the gluon and quark virtualities
(an attempt to take into account that the QCD
coupling is renormalized not only by the vector particle propagator,
but also by the quark-gluon vertex and the quark-propagator),
and to the choice of the renormalization scale
according to the Brodsky-Lepage-Mackenzie (BLM) procedure \cite{BrL83}
the essence of which is that all vacuum-polarization 
effects from the QCD $\beta$ function should be resummed into the
running coupling constant 
(the NLO coefficient of $T_H$ becomes $n_f$  (i.e. $\beta_0$)
independent).

\begin{figure}
\centerline{\epsfig{file=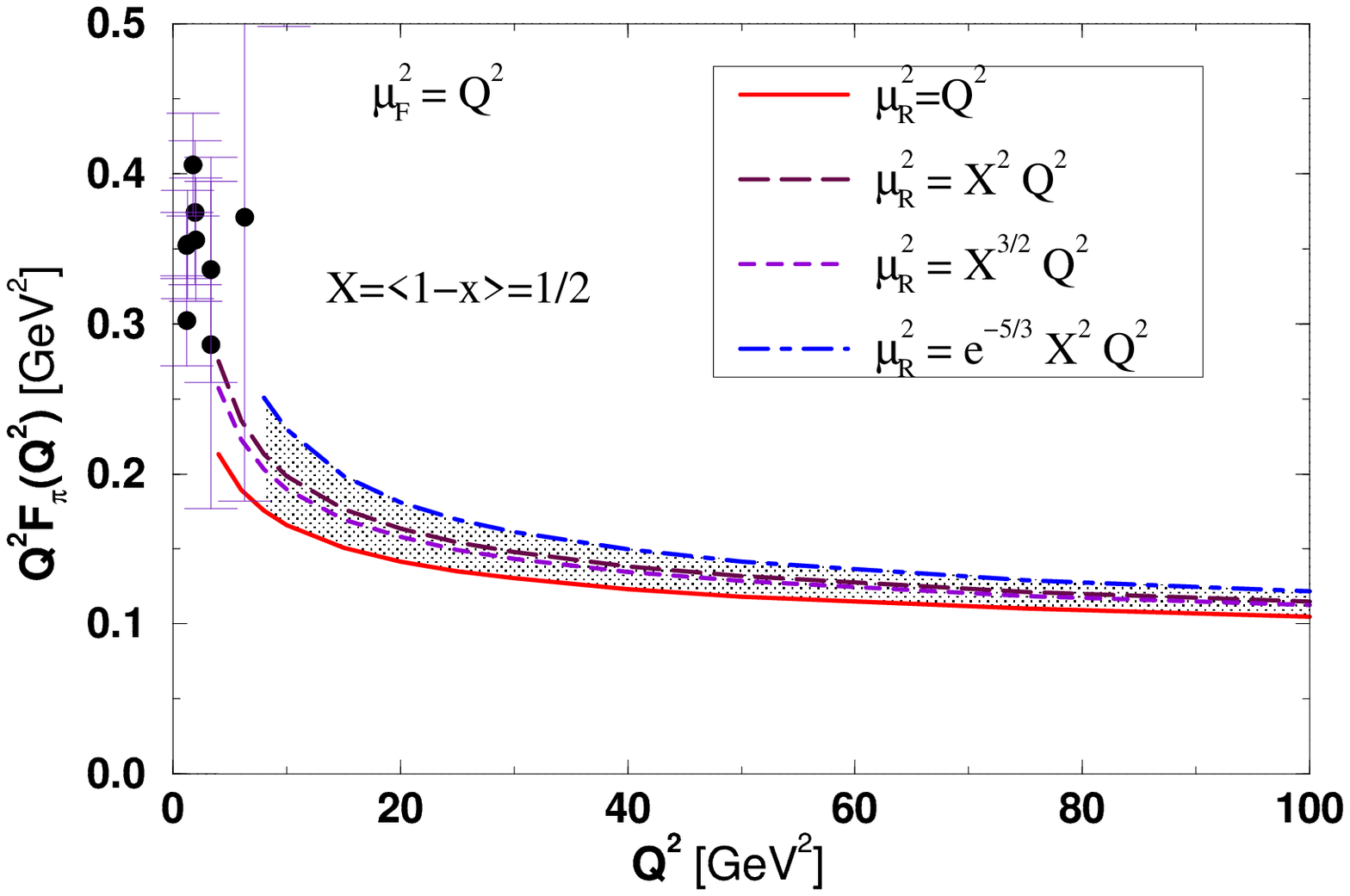,height=5.3cm,width=7.5cm,silent=}}
\caption{NLO prediction for $Q^2 F_{\pi}(Q^2)$ obtained
  using the $\phi_{as}(x,\nIR)$ distribution amplitude and
  the choices of $\nUV$ given by Eqs. \protect\req{eq:nRQ},
  \protect\req{eq:aQ2}, and \protect\req{eq:a}, 
  while $\nIR=Q^2$ and $\left< \x \right>=0.5$.
  The shaded area denotes the theoretical uncertainty introduced
  by the renormalization scale ambiguity.
  }
\label{f:ASbA}
\end{figure}

A glance at Eq. \req{eq:piffcf}, 
where the coupling constant $\alpha_S(\nUV)$ 
appears under the integral sign, 
reveals that the choice \req{eq:axyQ2}
leads immediately to the problem
if the usual one-loop formula 
for the effective QCD running coupling constant is employed.
Namely, regardless of how large $Q^2$ is,
the integrations over $x$ and $y$ 
allow $\alpha_S(\nUV)$ to be evaluated at low momenta
i.e., in the region where usual one-loop formula is not
a good representation of the effective QCD coupling.
There are number of proposals \cite{alphaSmod,alphaSmodn} 
for the form of the coupling constant
$\alpha_S(\nUV)$ for small $\nUV$, 
but its implementation in this calculation
demands the more refined treatment.\cite{ourps}
Alternatively, one can choose $\nUV$ to be an effective constant
\Beq
      \nUV=\left< \nUV \right>
          = \left< a(x,y) \right> \, Q^2
          = a \, Q^2 \, .
\label{eq:aQ2}
\Eeq

Hence, the expressions \req{eq:axy} get replaced by their respective
averages
\Beq
    a \in
         \{
{\left < \x \right >}^{2},
{\left < \x \right >}^{3/2},
e^{-5/3} {\left < \x \right >}^{2}
         \} \, .
\label{eq:a}
\Eeq
Now, the key quantity in the above expressions is 
$\left< \x \right>$, the average value of the momentum fraction.
Owing to the fact that $\phi_{as}(x,\nIR)$ 
is centered around the value $x=0.5$, 
the simplest choice is
$\left< \x \right>=0.5$.

\begin{figure}
\centerline{\epsfig{file=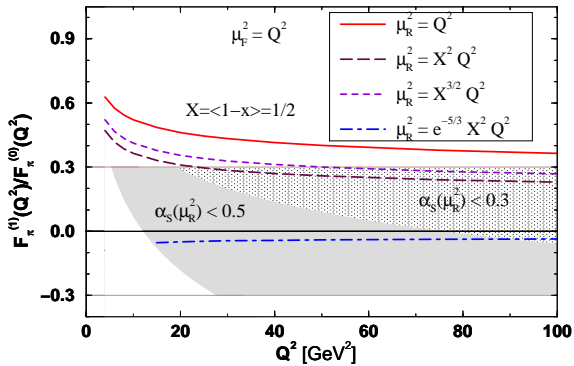,height=5.3cm,width=7.5cm,silent=}}
\caption{The ratio $F_{\pi}^{(1)}(Q^2)/F_{\pi}^{(0)}(Q^2)$
  obtained using the same DA and the same choices for
  $\nUV$ and $\nIR$ scales as in Fig. \protect\ref{f:ASbA} 
  The shaded area denotes the region of predictions which 
  corresponds to 
  $| F_{\pi}^{(1)}(Q^2)/F_{\pi}^{(0)}(Q^2) |<30\% $, and
  $\alpha_S(\nUV)\!<\!0.5$ ($\alpha_S(\nUV)\!<\!0.3$).
  }
\label{f:ratioASbA}
\end{figure}

Numerical results of our complete NLO QCD calculation 
for $Q^2 F_{\pi}(Q^2)$,
obtained using the $\phi_{as}(x,\nIR)$ distribution, 
with $\nIR=Q^2$ and different choices for the
renormalization scale $\nUV$ given by \req{eq:nRQ} 
\req{eq:aQ2}, \req{eq:a}, and $\left< \x \right>=1/2$,
are displayed in Fig. \ref{f:ASbA}
(in our calculation we take $\Lambda_{\overline{MS}}=0.2$
and $f_{\pi}=0.131$ GeV).
The ratio of the NLO to the LO contribution to 
$F_{\pi}(Q^2)$, i.e., $F_{\pi}^{(1)}(Q^2)/F_{\pi}^{(0)}(Q^2)$, 
as a useful measure of the importance of the NLO corrections,
is plotted as a function of $Q^2$ in Fig. \ref{f:ratioASbA}.

One notices that
the total NLO perturbative prediction for $Q^2 F_{\pi}(Q^2)$ 
is somewhat below the trend indicated 
by the presently available experimental data,
but what alarms us and 
seems to question the self-consistency of the PQCD approach 
is the fact that
the ratio $F_{\pi}^{(1)}/F_{\pi}^{(0)}$ 
corresponding to the often encountered choice $\nUV=Q^2$
(solid line) is rather high: 
$F_{\pi}^{(1)}/F_{\pi}^{(0)} \leq 30\%$ 
is not reached
until $Q^2 \approx 500$ GeV$^2$.
The answer to this problem lies in the previously stated inappropriateness
of the choice $\nUV=Q^2$. 
Namely, owing to the partitioning of the overall momentum
transfer $Q^2$ among the particles in the parton subprocess,
the essential virtualities of the particles are smaller than $Q^2$,
so that the ``physical'' renormalization scale, better suited for the
process of interest, is inevitably lower than $Q^2$.
It follows from Fig. \ref{f:ratioASbA}
that by choosing the renormalization scale
determined by the dynamics of the pion rescattering process, the size of
the NLO corrections is significantly reduced and reliable predictions are
obtained  at considerably lower values of $Q^2$, namely, for $Q^2<100$ GeV$^2$.

The total NLO prediction for
$Q^2 F_{\pi}(Q^2)$ 
depends weakly on the choice of $\nUV$.
This is a reflection of the stabilizing effect that the inclusion of the
NLO corrections has on the LO predictions.
Let us explore this point more closely.
By taking $\nUV$ and $\nIR$ to be effective constants,
the LO and NLO contributions to the pion form factor
obtained using $\phi_{as}(x,\nIR)$ distribution
amount to
\Beqa
  Q^2 F_{\pi}^{(0)}(Q^2,\nUV,\nIR) & = &  
            8 \, \pi \, f_{\pi}^2 \,  
                \alpha_S( \nUV ) \, ,
		 \label{eq:Q2Fpi0}\\
  Q^2 F_{\pi}^{(1)}(Q^2,\nUV,\nIR) 
               & = & 8 \,  f_{\pi}^2  \,
                \alpha_S^2( \nUV ) \,
                \left[
		   \frac{\beta_0}{4} 
                   (\ln \frac{\nUV}{Q^2} + \frac{14}{3}) 
		  -3.92  \right] \, ,
		 \label{eq:Q2Fpi1}
\label{eq:res}
\Eeqa
where the NLO contribution coming from the NLO evolution
of the DA is neglected.
As it is seen from \req{eq:Q2Fpi0},
all of the $\mu_R$ dependence of the LO result $Q^2 F_{\pi}^{(0)}$
is contained in the strong coupling constant
$\alpha_S(\nUV)$.
Thus, 
as $\mu_R$ decreases the LO result increases, and
it increases without bound.
In contrast to the LO,
the NLO contribution
$Q^2 F_{\pi}^{(1)}$, as evident from the explicit expression
given in \req{eq:Q2Fpi1}, decreases (becomes more negative)
with decreasing $\mu_R$.
Upon adding up the LO and NLO contributions, 
we find that the full NLO result,
as a function of $\mu_R$ stabilizes and reaches
a maximum value for $\nUV=\mu_{extreme}^2 \approx Q^2/18$. 
If we take that the renormalization scale continuously changes 
in the interval defined by $a \in [1/18,1]$
the curves representing the NLO predictions for
$Q^2 F_{\pi}(Q^2)$ 
fill out the shaded region in Fig. \ref{f:ASbA}, 
and this shaded region essentially determines 
the scale ambiguity related theoretical uncertainty of  
the NLO calculation.

\section{Towards resolving the renormalization scale ambiguity
         problem}

The optimization of the scale and scheme choice according to some
sensible criteria remains 
an important task for the application of PQCD.
Several scale-setting procedures were proposed in the literature:
the principle of fastest apparent convergence 
(FAC) \cite{FAC}
($F_{\pi}^{(1)}(Q^2,\mu_{FAC}^2)=0$),
%($\nUV$ is determined by the requirement that NLO correction
%vanishes),
the principle of minimal sensitivity 
(PMS) \cite{PMS}
($\mu_{PMS}^2=\mu_{extreme}^2$), 
%($\nUV$ is chosen at the stationary point of the NLO prediction
%that is in our case at $\nUV=\mu_{extreme}^2$),
and the BLM method.
The application of those methods can give 
strikingly different results in some calculations.\cite{KramL91}

As it is known, the relations between physical observables must
be independent of renormalization scale and scheme
conventions to any fixed order of perturbation theory.
It was argued \cite{BroL95} that
applying the BLM scale-fixing to perturbative predictions
of two observables
in, for example, $\overline{MS}$ scheme
and then algebraically eliminating 
$\alpha_{\overline{MS}}$ one can relate 
any perturbatively calculable observables without scale and scheme
ambiguity, where the choice of BLM scale ensures
that the resulting ``commensurate scale relation'' (CSR)
is independent of the choice of the intermediate renormalization
scheme.
Following this approach, 
in paper by Brodsky {\em et al.}\cite{BroJ98}
the exclusive hadronic amplitudes
were calculated in
$\alpha_V$ scheme, 
in which the effective coupling $\alpha_V(\mu^2)$ is defined
from the heavy-quark potential $V(\mu^2)$.

\begin{table}
\caption{NLO PQCD results for the pion form factor,
         $Q^2 F_{\pi}(Q^2)$ in the $\alpha_V$ scheme.}
\centerline{
\begin{tabular}{|c|c|c|c|} 
\hline 
$Q^2$ $[\mbox{GeV}^{2}] $& $\alpha_V(\mu_V^2)$ & 
$F_{\pi}^{(1)}(Q^2)/F_{\pi}^{(0)}(Q^2)$ $[ \% ]$ & 
$Q^2 F_{\pi}(Q^2)$ $[\mbox{GeV}^{2}]$ 
 \\[.05cm] \hline  
 20 & 0.434 & -26.5 & 0.138 \\
% 50 & 0.338 & -23.6 & 0.127 \\
% 75 & 0.308 & -18.8 & 0.108 \\
100 & 0.289 & -17.7 & 0.103 \\
\hline
\end{tabular}
}

\label{t:Vsch}
\end{table}

It follows from \req{eq:Q2Fpi1} that 
the $\beta_0$ dependent term vanishes for 
$\nUV=\left< \mu_{BLM}^2 \right> =e^{-14/3} Q^2$.
Using the scale-fixed relation \cite{BroJ98}
between $\alpha_V(\mu_V^2)$ and 
$\alpha_{\overline{MS}}(\mu_{\overline{MS}}^2)
\equiv \alpha_S(\nUV)$,
and Eqs. (\ref{eq:Q2Fpi0}-\ref{eq:Q2Fpi1}) 
one obtains the NLO prediction for the pion form factor
in $\alpha_V$ scheme \cite{BroJ98}
\Beq
  Q^2 F_{\pi}(Q^2,\mu_V^2)  =   
            8 \, \pi \, f_{\pi}^2 \,  
                \alpha_V(\mu_V^2) 
      \left( 1 -1.92
      \frac{\alpha_V(\mu_V^2)}{\pi} \,
             \right) \, ,
\label{eq:resV}
\Eeq
where $\mu_V^2=e^{5/3} \nUV=e^{-3} Q^2 \approx Q^2/20$.
Considering the energy region we are interested in,
for the purpose of this calculation  we 
approximate $\alpha_V$ with usual one-loop
expression. 
Thus obtained numerical predictions are given in
Table \ref{t:Vsch}.
Considering the size of the 
$F_{\pi}^{(1)}/F_{\pi}^{(0)}$ ratio ($\approx -20\%$)
and the size of the effective coupling, we find that 
NLO PQCD predictions for the pion form factor
obtained in $\alpha_V$ scheme can be considered reliable
for $Q^2$ below $100$ GeV$^2$ 
i.e., already for momentum transfer
$Q$ of the order of $5$ -- $9$ GeV. 

\section{Conclusions}

We have shown that,
regarding the size of the radiative corrections, 
the sHSA can be consistently applied to the calculation
of the pion electromagnetic form factor already  
for momentum transfer $Q$ of the order of $5$ -- $10$ GeV.
In order to improve this PQCD based prediction it is necessary
to obtain and apply the proper form of the QCD coupling in the  
low-momentum regime
as well as to investigate the corrections (present in the few GeV
region) introduced by  the mHSA.\cite{SteS99} 
Reliable experimental data and inclusion of the soft contributions
should then enable complete confrontation between theory and
experiment.

\section*{Acknowledgments}

  This work was supported by the Ministry of Science and Technology
  of the Republic of Croatia under Contract No. 00980102.
  One of us (K.P.) would like to thank Anatoly Radyushkin
  and Carl Carlson for the opportunity to participate in this
  stimulating workshop.

\end{document}